\documentclass[twocolumn,aps,prb,amsfonts]{revtex4}
\usepackage{graphicx}
\usepackage{psfrag, color}
\usepackage{verbatim}

\begin{document}

\title{Scattering Mechanisms in a High Mobility Low Density Carbon-Doped (100) GaAs Two-Dimensional Hole System}

\author{J. D. Watson$^1$, S. Mondal$^1$, G. A. Cs\'{a}thy$^1$, M. J. Manfra$^{1,2,3,}$\footnote[1]{mmanfra@purdue.edu}, E. H. Hwang$^4$, S. Das Sarma$^4$, L. N. Pfeiffer$^5$, K. W. West$^5$}

\affiliation{${}^1$ Department of Physics and Birck Nanotechnology Center\\
${}^2$ School of Electrical and Computer Engineering\\
${}^3$School of Materials Engineering\\
Purdue University,
West Lafayette, IN 47907\\
${}^4$Condensed Matter Theory Center and Department of Physics\\ University of Maryland, College Park, MD, 20742\\
${}^5$Department of Electrical Engineering\\
Princeton University,
Princeton, NJ 08544\\}

\begin{abstract}
We report on a systematic study of the density dependence of mobility
in a low-density Carbon-doped (100) GaAs two-dimensional hole system
(2DHS).  At T=50\,mK, a mobility of 2.6 $\times$ 10$^{6}$ cm$^{2}$/Vs
at a density p=6.2 $\times$ 10$^{10}$ cm$^{-2}$ was measured.  This is
the highest mobility reported for a 2DHS to date. Using a back-gated
sample geometry, the density dependence of mobility was studied from
2.8 $\times$ 10$^{10}$ cm$^{-2}$ to 1 $\times$ 10$^{11}$ cm$^{-2}$.
The mobility vs. density cannot be fit to a power law dependence of
the form $\mu$ $\sim$ p$^{\alpha}$ using a single exponent $\alpha$.
Our data indicate a continuous evolution of the power law with
$\alpha$ ranging from $\sim$ 0.7 at high density and increasing to
$\sim$ 1.7 at the lowest densities measured.  Calculations specific to
our structure indicate a crossover of the dominant scattering
mechanism from uniform background impurity scattering at high density
to remote ionized impurity scattering at low densities. 
This is the first observation of a carrier density-induced
transition from background impurity dominated to remote dopant
dominated transport in a single sample. 
\end{abstract}

\maketitle
The two-dimensional hole system \,(2DHS) offers an attractive platform
for the study of strong carrier interactions parameterized by $r_s$ :
the ratio  of the Coulomb energy to the Fermi energy. $r_s = E_c/E_f
\propto m^*/\sqrt{p}$, where $p$ is the hole density and $m^*$ is the
effective mass.  Recent developments in the growth of Carbon-doped
(100) GaAs heterostructures by molecular beam epitaxy (MBE) have
resulted in 2DHSs of unprecedented quality \cite{gerl2005carbon}${}^,$
\cite{manfra2005high}.  Such structures have been utilized in the
study the metal-to-insulator transition (MIT)
\cite{manfra2007transport}, fractional quantum Hall physics in the 2nd
Landau level (LL)\cite{kumar2010particle}, spin-orbit coupling in
Aharonov Bohm rings \cite{grbic2007aharonov} and charge density wave
formation in partially filled
LL's\cite{koduvayur2010effect}${}^,$\cite{manfra2007impact}. These
initial experiments and the prospect of studying strong correlations
in the presence of tunable spin-orbit coupling provide strong
motivation to understand the scattering processes presently limiting
mobility in the highest quality samples.  Here we present mobility
vs.\ density data on an unprecedently high mobilty 2DHS.  One of the
most exciting avenues for future research is the investigation of
ultra-low density 2DHSs at very large $r_s$.  Thus, our data and
calculations will inform the design of new hole heterostructures of
ever increasing quality. 

Carbon doping\cite{manfra2005high} of 2DHSs offers advantages over the
more commonly used acceptor dopants Beryllium and Silicon.  Carbon
diffuses and surface segregates much less at typical MBE
growth temperatures (T$\sim$630 $^{\circ}$C) than
Beryllium\cite{schubertbook}.  Additionally Carbon can be
incorporated as an acceptor on multiple crystallographic orientations,
including on the high-symmetry (100) face of GaAs.  Silicon can also
act as an acceptor to produce high quality 2DHSs but so far high
mobility ($\mu \sim 10^6$ cm$^2$/Vs) Silicon-doped 2DHSs have only
been realized on (311)A face\cite{heremans1994mobility}.  The (311)A
face has a well known mobility anisotropy between the [$\bar{2}33$]
and [$01\bar{1}$] directions \cite{heremans1994mobility} whereas
Carbon-doped structures on the (100) face have a significantly lower
anisotropy between the [$011$] and [$0\bar{1}1$] directions
\cite{manfra2005high}.  Furthermore, the high symmetry of the (100)
orientation dramatically alters the nature of spin-orbit interactions
in 2DHSs as compared to quantization along the (311)A direction.
Indeed, further experimental work is needed to fully exploit the
potential benefits of Carbon-doped (100) 2DHSs. 

\begin{figure}[!t]
\includegraphics[width=3.25in]{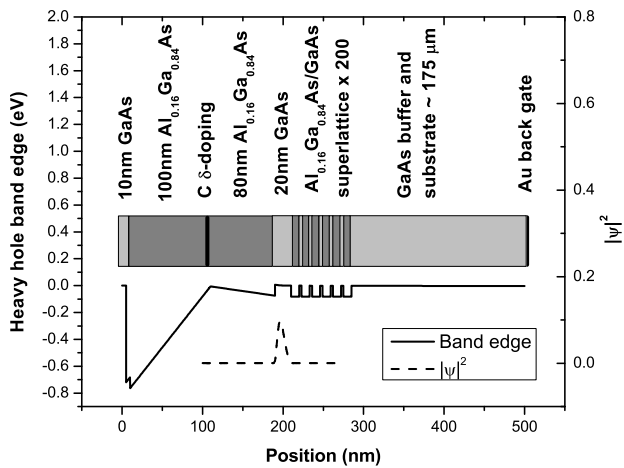}
\caption{Schematic of the device structure used in our experiments.}
\label{QW}
\end{figure}

\begin{figure}[!b]
\includegraphics[width=3in]{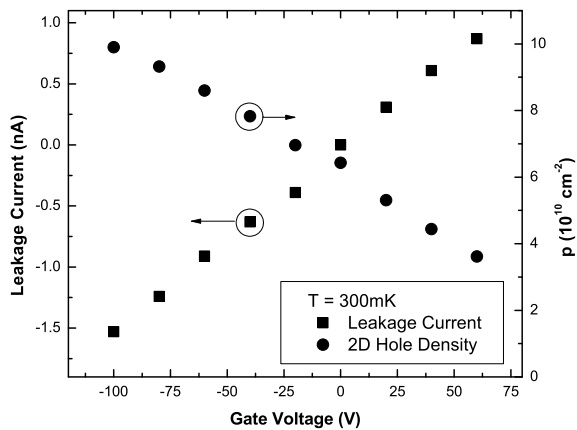}
\caption{Effect of gate voltage on carrier density and leakage current at T = 300 mK.}
\label{Vg}
\end{figure}

Our sample consists of a 20\,{}nm Al$_{0.16}$Ga$_{0.84}$As/GaAs/Al$_{0.16}$Ga$_{0.84}$As quantum well asymmetrically modulation doped with Carbon at a density of 1 $\times$ 10$^{12}$ cm$^{-2}$ above the quantum well at a setback of 80 nm.  FIG.\,{}\,{}\ref{QW} shows a sketch of the device along with the numerically calculated \cite{nextnano} band structure and heavy hole ground state wavefunction (normalized to unity).  For simplicity in simulation the superlattice and buffer regions are truncated.
In order to modulate the density in the quantum well, we utilized a back-gate geometry.  The sample was first thinned to approximately 150 $\mu$m and then cleaved into a Hall bar approximately 2 mm $\times$ 9 mm.  Ohmic contacts consist of In/Zn dots positioned approximately 1 mm apart along the length of the Hall bar and annealed at T = 430 $^{\circ}$C.  The hall bar was subsequently fixed to a gold backgate evaporated on an undoped GaAs substrate.  The carrier density was measured from minima in the longitudinal magnetoresistance, and the conductivity was obtained from four-terminal zero field measurements using standard lock-in techniques.  As shown in FIG. \ref{Vg}, the 2DHS density depended linearly on voltage over the range measured.  Modeling the structure as a parallel plate capacitor with one plate being the 2DHS and the other being the backgate we estimate the gate to be situated 175 $\mu$m from the well.  The peak mobility $\mu$ at low temperature (T = 50 mK) was measured to be 2.6 $\times$ 10$^{6}$ cm$^{2}$/Vs at a density of 6.2 $\times$ 10$^{10}$ cm$^{-2}$ in an as-grown sample.

\begin{figure}[!t]
\includegraphics[width=3in]{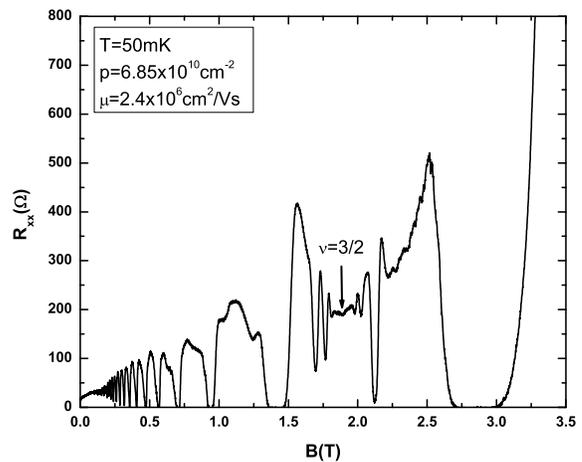}
\caption{Low field magnetoresistance of the backgated sample at T = 50 mK.}
\label{MR}
\end{figure}

FIG. \ref{Vg} also shows the leakage current as a function of the gate voltage.  The linear dependence of the leakage current on the gate voltage and its small magnitude ($<$ 1.5 nA as compared to an excitation current of 50 nA) suggest that the observed leakage represents parasitic current passing through the measurement circuit and not hard breakdown in the GaAs.  In addition, the linear dependence of the density on the gate voltage also suggests that sharp breakdown did not occur.  FIG. \ref{MR} shows a representative trace of the magnetoresistance at T = 50 mK.  The deep minima in the fractional quantum Hall states around filling factor $\nu = \frac{3}{2}$ illustrate the high quality of the processed sample.  We note that this sample has also been studied at ultra-low temperatures (T $\leq$ 10 mK) in which the first evidence of a fully formed fractional quantum Hall  state at $\nu = 8/3$ in the 2nd Landau level in a 2DHS was observed \cite{kumar2010particle}.

\begin{figure}[t]
\includegraphics[width=3in]{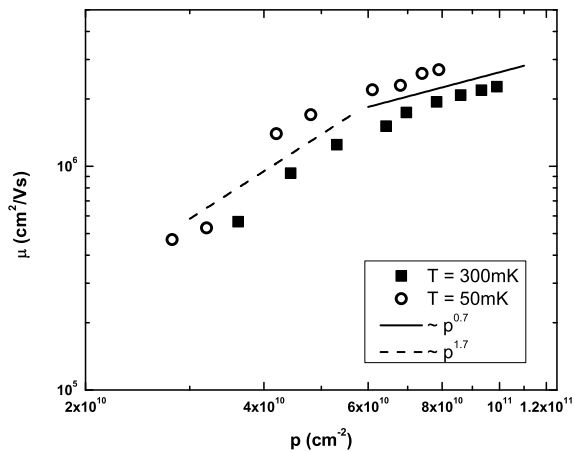}
\caption{Mobility as a function of the density at T = 300 mK (squares) and T = 50 mK (open circles).  Straight lines are guides to the eye to the 300 mK data to illustrate 0.7 and 1.7 power laws.}
\label{n_v_mu}
\end{figure}

In order to examine the scattering mechanisms limiting mobility in our system, we measured the dependence of the mobility on the 2D hole density modulated by the backgate as shown in FIG. \ref{n_v_mu}.  As can be clearly seen on this log-log plot the data points do not fall on a straight line as would be expected for a single dominant scattering mechanism. The mobility vs. density cannot be fit to a power law dependence of the form $\mu$ $\sim$ p$^{\alpha}$ using a single exponent $\alpha$.  Our data indicate a continuous evolution of the power law with $\alpha$ ranging from $\sim$ 0.7 at high density and increasing to $\sim$1.7 at the lowest densities measured. Thus the data indicate the presence of multiple dominant scattering mechanisms over the range of density tested.  Indeed at the lowest densities measured, the mobility decreases rapidly indicating that the system will eventually approach a finite density MIT\cite{manfra2007transport}.  We emphasize, however, that $k_Fl$, the product of the Fermi wavevector and the mean free path, remains larger than 50 over the entire range of density tested.  It can be seen at high density that the mobility follows a power law behavior $\mu \propto p^{0.7}$ which is indicative of uniformly distributed charged background impurity (BI) scattering \cite{hwang2008limit}${}^,$ \cite{umansky1997extremely} in 2D carrier systems.  
\begin{figure}[h!]
\includegraphics[width=3in]{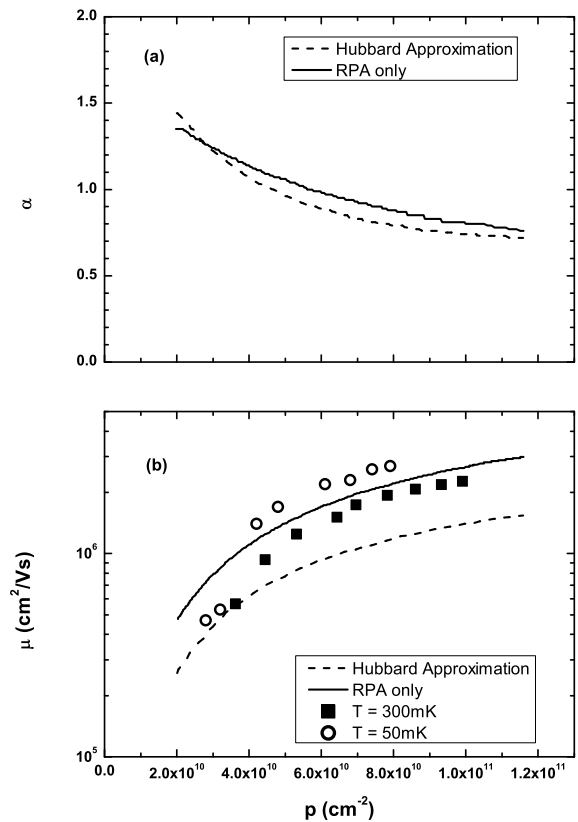}
\caption{(a) Theoretical density dependence of the exponent $\alpha$ in $\mu \propto p^{\alpha}$.  (b) Comparison of experimental mobility data and theoretical results.  Solid line represents RPA-Boltzmann calculation and dashed line represents RPA-Boltzmann calculation with the Hubbard approximation in both plots.}
\label{theory}
\end{figure}
However, the power law continuously shifts towards higher $\alpha$ of
approximately 1.7 at low density. Typically, an exponent $\sim
1.5-1.7$ is taken as an indication of the dominance of remote ionized
(RI) impurity scattering\cite{hwang2008limit}$^,$\cite{lee1983low}
originating in the remote doping layer in 2D systems.  We emphasize
that it is unusual that such a crossover behavior can be seen in a
single sample while remaining in the high mobility (or equivalently
large $k_Fl$) regime. It is interesting to note that our setback
distance is 80 nm, a distance at which remote ionized impurity
scattering usually makes a relatively minor contribution to the total
scattering in samples with density $\gtrsim 10^{11}$ cm$^{-2}$.
Nevertheless, it is clear that for the material parameters of our
structure it dominates scattering at lower density.  A similar
transition to remote ionized impurity scattering in low density 2D
electron samples was report by Jiang et al. \cite{jiang1988threshold},
but in these samples $k_Fl$ was substantially lower and the samples
were approaching a conduction threshold. 
Unusual transport behavior was also recently reported in the 2DHS in
an undoped electron-hole bilayer \cite{seamons2007undoped} and
subsequently explained by carrier inhomogeneities resulting from
strong carrier-carrier interactions and non-linear screening
\cite{hwang2008transport}.  It could be argued that T = 300 mK data is
only marginally outside the range where phonons could be playing some
role, but the fact that our T = 50 mK data displays the same
functional dependence seems to rule out any critical scattering
contributions from phonons.  Interface roughness (IR) scattering
should also be considered, but significant IR scattering should also
manifest itself in a significant mobility anisotropy
\cite{umansky1997extremely} which was not observed in our sample.
Finally, scattering between the light and heavy hole
bands\cite{gerl2005carbon} as well as changes in the effective mass
\cite{lu2008cyclotron} have also been reported.  However, both of
these effects should decrease as the density is lowered and the Fermi
level moves towards the top of the valence band and away from light
hole band. 

To explain our experimental observation we calculated the Coulomb
scattering rate due to background charged impurity scattering and
remote charged impurity scattering using a Boltzmann transport
method\cite{dassarma1999charged}${}^,$ \cite{dassarma2000calculated}.
Screening was taken into account using the random phase approximation
(RPA).  The calculation was performed both with and without the
inclusion of correlation effects via the Hubbard approximation.  The
analysis assumed a 3D background impurity concentration of
$n_{i3D}=3\times 10^{13}$ cm$^{-3}$, dopant setback from the center of
the quantum well $d = 90$ nm, hole effective mass $m^*=0.4m_e$ where
$m_e$ is the free electron mass, quantum well width $a = 20$ nm, and a
remote ionized impurity concentration $n_i = 2\times 10^{11}$
cm$^{-2}$.  FIG. \ref{theory}(a) shows a crossover of the mobility
exponent changing from $\alpha \sim 0.7-0.8$ to $\alpha \sim 1.5-1.7$
which has the same qualitative behavior as seen in the data in
FIG. \ref{n_v_mu}.  The results of the calculations are compared with
the experimental data in FIG. \ref{theory}(b).  Qualitatively, as the
hole density is lowered, screening of the remote dopants by the 2D
hole gas becomes less effective, and the dominant scattering mechanism
transitions from being dominated by uniform background impurities to
being dominated by remote dopants.  Such a transition point in the
density is governed entirely by $n_{i3D}$, $d$, and $n_i$, and thus it
is not surprising that this is the first time (to our knowledge) that
such a transition has been observed in a single sample within the high
mobility regime. 

To understand the transition in the transport mechanism observed in
our data, it is important to realize\cite{hwang2008transport} that the
impurity scattering 
strength in the transport theory carries the form-factor $\exp(-2k_Fd)$
at low temperatures where $2k_F$ is the typical momentum transfer for
resistive scattering by impurities, and $d$ is the typical separation of
the impurities from the 2D carrier layer.  
Since $k_F \sim p^{1/2}$, it is
clear that lowering the carrier density would lead to stronger
scattering by remote impurities which is exponentially suppressed at
higher values of $k_Fd$.  
For a screened Coulomb potential with two impurity contributions (remote
and background charged impurity) we can
derive the approximate qualitative formula for the mobility 
\begin{equation}
\mu \propto \frac{(k_Fd)^3q_{TF}}{n_i + A n_{i3D} (k_Fd)^3
  q_{TF}/(2k_F + q_{TF})^2}
\end{equation}
where $A$ is a density independent constant and $q_{TF}=2/a_B$ is the
Thomas Fermi screening wave vector with the effective Bohr radius
$a_B=\hbar^2/m^*e^2$.  
In the high-density limit, $k_Fd \gg 1$ the mobility is
proportional to the square of the sum of two wave vectors, i.e., $\mu
\propto (2k_F + q_{TF})^2$. 
However, as $k_F d$ (or, density) decreases the mobility behaves approximately $\mu
\propto (k_Fd)^3$.
Thus, as long as strong localization does not
set in, which is the usual situation for lower mobility
samples\cite{jiang1988threshold}, 
lowering carrier density should always lead to a continuous increase
of the exponent $\alpha$ as scattering from the remote dopant
impurities becomes important.  This is exactly the theoretical
behavior predicted in the theory (Fig.~\ref{n_v_mu}), and experimentally
observed in our extremely high-mobility hole sample.

In conclusion, we measured the density dependence of mobility in a
very high quality 2DHS.  The 50 mK mobility was found to be 2.6
$\times$ 10$^6$ cm$^2$/Vs at a density of 6.2 $\times$ 10$^{10}$
cm$^{-2}$ in a pristine sample.  The mobility appears to be limited by
background charged impurity scattering at high density but in the low
density regime is a stronger function of the density indicating an
increasingly important scattering contribution from remote impurities.
From this data, we can surmise that increased 2DHS mobility can be
realized at low density by significantly increasing the spacer
thickness.  Such experiments are currently underway. 
Our work also demonstrates that in samples of sufficiently high quality,
where the 2D MIT transition is pushed down to very low carrier
densities, the theoretically predicted continuous transition from background impurity scattering limited transport to remote dopant
scattering limited transport can be quantitatively verified by
decreasing the carrier density in a single sample. 
%

\section*{Acknowledgment}
\vspace*{-10pt}
JDW is supported by a Sandia Laboratories/Purdue University
Excellence in Science and Engineering Fellowship.
MJM acknowledges support from the Miller Family Foundation. 
The work by Maryland (EH and SDS) is supported by Microsoft Q.  GAC acknowledges support from the NSF DMR-0907172.
%

\begin{thebibliography}{50}
\bibitem{gerl2005carbon} C. Gerl, S. Schmult, H.-P. Tranitz, C. Mitzkus, and W. Wegscheider, Appl. Phys. Lett. {\bf 86}, 252105 (2005).
%
\bibitem{manfra2005high} M. J. Manfra, L. N. Pfeiffer, K. W. West, R. de Picciotto, and K. W. Baldwin, Appl. Phys. Lett. {\bf 86}, 162106 (2005).
%
\bibitem{manfra2007transport} M. J. Manfra, E. H. Hwang, S. Das Sarma, L. N. Pfeiffer, K. W. West, and A. M. Sergent, Phys. Rev. Lett. {\bf 99}, 236402 (2007).
%
\bibitem{kumar2010particle} A. Kumar, N. Samkharadze, M. J. Manfra, G. A. Cs\'{a}thy, L. N. Pfeiffer, K. W. West, arXiv:1007.1504v1 (2010).
%
\bibitem{grbic2007aharonov} B. Grbi\'{c}, R. Leturcq, T. Ihn, K. Ensslin, D. Reuter, and A. D. Wieck, Phys. Rev. Lett. {\bf 99}, 176803 (2007).
%
\bibitem{koduvayur2010effect} S. P. Koduvayur, Y. Lyanda-Geller, S. Khlebnikov, G. Cs\'{a}thy, M. J. Manfra, L. N. Pfeiffer, K. W. West, and L. P. Rokhinson, Phys. Rev. Lett. {\bf 106}, 016804 (2011).
%
\bibitem{manfra2007impact} M. J. Manfra, R. de Picciotto, Z. Jiang, S. H. Simon, L. N. Pfeiffer, K. W. West, and A. M. Sergent, Phys. Rev. Lett. {\bf 98}, 206804 (2007).
%
\bibitem{schubertbook} E. F. Schubert, \textit{Doping in III-V Semiconductors} (Cambridge University Press, Cambridge, 1993).
%
\bibitem{heremans1994mobility} J. J. Heremans, M. B. Santos, K. Hirakawa, and M. Shayegan, J. Appl. Phys. {\bf 76}, 1980 (1994).
%
\bibitem{nextnano} Nextnano3 simulator \copyright 1999-2008 Walter Schottky Institute.  http://www.nextnano.de/nextnano3/index.htm.
%
\bibitem{hwang2008limit} E. H. Hwang and S. Das Sarma, Phys. Rev. B {\bf 77}, 235437 (2008).
%
\bibitem{umansky1997extremely} V. Umansky, R. de Picciotto, and M. Heiblum, Appl. Phys. Lett. {\bf 71}, 683 (1997).
%
\bibitem{lee1983low} K. Lee, M. S. Shur, T. J. Drummond, and H. Morko\c{c}, J. Appl. Phys. {\bf 54}, 6432 (1983).
%
\bibitem{jiang1988threshold} C. Jiang, D. C. Tsui, and G. Weimann, Appl. Phys. Lett. {\bf 53}, 1533 (1988).
%
\bibitem{seamons2007undoped} J. A. Seamons, D. R. Tibbetts, J. L. Reno, and M. P. Lilly, Appl. Phys. Lett. {\bf 90}, 052103 (2007).
%
\bibitem{hwang2008transport} E. H. Hwang and S. Das Sarma, Phys. Rev. B {\bf 78}, 075430 (2008).
%
\bibitem{lu2008cyclotron} T. M. Lu, Z. F. Li, D. C. Tsui, M. J. Manfra, L. N. Pfeiffer, and K. W. West, Appl. Phys. Lett. {\bf 92}, 012109 (2008).
%
\bibitem{dassarma1999charged} S. Das Sarma and E. H. Hwang, Phys. Rev. Lett. {\bf 83}, 164 (1999).
%
\bibitem{dassarma2000calculated} S. Das Sarma and E. H. Hwang, Phys. Rev. B {\bf 61}, 7838 (2000).
%
\end{thebibliography}

\vspace*{-10pt}



\end{document}